\documentclass[aps,prx,twocolumn,superscriptaddress]{revtex4}
\usepackage{amsmath}
\usepackage{dcolumn}
\usepackage{amssymb}
\usepackage{amsfonts}
\usepackage{graphicx}
\usepackage[T1]{fontenc}
\usepackage{indentfirst}
\usepackage{color}
\usepackage{bbm}
\usepackage{amsbsy}
\usepackage{ulem}
\usepackage{cancel}
\usepackage[dvipsnames]{xcolor}

\begin{document}
\title{ 2$k_F$ instability and chiral spin density wave at the 1/9 magnetization plateau in the kagome antiferromagnets}
\author{Tanja \DJ uri\'c}
\affiliation{School of Physical and Mathematical Sciences, Nanyang Technological University, 21 Nanyang Link, Singapore 637371}
\author{Pinaki Sengupta}
\affiliation{School of Physical and Mathematical Sciences, Nanyang Technological University, 21 Nanyang Link, Singapore 637371}
\affiliation{Department of Physics, Boston University, 590 Commonwealth Avenue, Boston, MA 02215, USA}
\date{\today}
\begin{abstract}
Kagome lattice antiferromagnets exhibit plethora of intriguing phases of matter. Particularly interesting state appears at the magnetic field-induced $1/9$ magnetization plateau observed in several recent experimental studies. The nature and exotic physical properties of the plateau however remain controversial due to an exceptional complexity of the state generated by geometrical frustration. Among candidate states recent studies found a $Z_3$ quantum spin liquid state, a valence bond crystal exhibiting an hourglass pattern and a valence bond crystal state with a $3\times 3$ periodicity and a windmill-shaped motif. Recent torque magnetometry measurements on YCOB single-crystal samples however indicate presence of Dirac-like spinons at $1/9$ magnetization plateau. We study properties of the plateau state using novel machine learning technique that combines variational Monte Carlo, symmetry enhanced neural network quantum states and flux insertion method. Our machine learning study reveals that the ground state at the $1/9$ plateau is a gapless $1\times 1$ chiral spin density wave caused by 2$k_F$ instability of the underlying composite Fermi liquid. The spin wave chirality results from the correlated spin order that reflects its nontrivial topology. 
\end{abstract}

\maketitle

\section{Introduction}
\label{sec:Introduction}

Numerous studies show that presence of geometric frustration in quantum magnets results in a wide variety of exotic phases of matter. Kagome lattice antiferromagnets  are paradigmatic examples of frustrated quantum magnets that can host various complex quantum states exhibiting unique properties such as long-range quantum entanglement, topological order and excitations with fractional quantum numbers \cite{Duric,Nishimoto,He,Fang,Zheng}. In the presence of an applied magnetic field, the magnetization exhibits a sequence of field induced plateaus. Particularly intriguing state appears at the $1/9$ magnetization plateau \cite{Nishimoto,He,Fang,Zheng}. The nature of the plateau  remains a matter of controversy since various studies using different methods lead to different conclusions about the properties of the ground state at the plateau. While density matrix renormalization group (DMRG) calculations \cite{Nishimoto} and recent variational Monte Carlo (VMC) studies \cite{He}, based on fermionic parton representation for spin operators and Gutzwiller projection, indicate that the plateau ground state is a $Z_3$ quantum spin liquid, infinite projected entangled pair states (iPEPS) calculations find a valence bond crystal ground state exhibiting an hourglass pattern \cite{Fang}. In addition, a separate variational calculation based on more general resonating valence bond (RVB) ansatz find a VBC state with a $3\times 3$ periodicity and a windmill-shaped motif \cite{Cheng}. Recent experimental findings based on torque magnetometry measurements on YCOB single-crystal samples, however, indicate presence of Dirac-like spinons at the 1/9 magnetization plateau that manifest as unconventional magnetic oscillations in magnetic torque \cite{Zheng}.  

In the past few years, symmetry enhanced neural network quantum states (NQS), and in particular group equivariant convolutional neural networks (GCNNs), have emerged as powerful ans\"atze that can describe complex states of matter \cite{Duric, Roth}. GCNNs combined with novel machine learning (ML) algorithms based on stochastic reconfiguration method \cite{Sorella1, Sorella2, Rende, Chen} allow high accuracy variational Monte Carlo (VMC) calculations. In this study we examine the nature of the $1/9$ magnetization plateau in kagome antiferromagnets by employing such novel ML techniques combined with GCNNs and flux insertion method. Contrary to previous studies our results reveal $2k_F$ instability \cite{Jian,Altshuler,Patel,Holder} of the underlying spinon Fermi surface resulting in a spin density wave ground state at the 1/9 magnetization plateau. Due to presence of $Z_3$ flux the state is best described by composite fermionic spinons composed of spinons with attached flux. In addition to long-range interactions mediated by the gauge field, there are also local interactions between such composite fermions that can cause pairing and 2$k_F$ density wave instability. 
Normally,  pairing instability is marginally relevant in a composite Fermi liquid. But in this case any pairing instability is suppressed due to presence of an emergent gauge boson~\cite{Jian}. 

In our calculations, the  presence of an emergent flux is confirmed by a finite chiral order parameter and the 2$k_F$ instability by the spin structure factor calculations. The spin structure factor can clearly identify 2$k_F$ scattering features caused by scattering close to the Fermi surface \cite{He2,Xu,Sheng} and 2$k_F$ instability that causes appearance of sharp peaks in the structure factor reflecting spin density wave ordering (SDW) with the wave vector corresponding to the position of the peaks \cite{Jian}. In addition our calculations of the spin components parallel to the applied field reveal $1\times1$ density wave ordering similar to the $1\times1$ charge density wave order found in kagome metals \cite{OBrien,Nishimoto2,Ferrari,Wen,Lee}. The $1\times 1$ chiral  spin density wave (CSDW) ground state maintains original translation symmetry of the kagome lattice and has spin modulation within the kagome lattice unit cell. Moreover results for the cylinder geometry show generalized Friedel oscillations \cite{Dora, White} due to presence of the SDW.

We also point out that the found SDW has "hot spots" situated at high symmetry points of the Fermi surface. The "hot spots" are locations on the Fermi surface where the density wave ordering is the most pronounced and where the nesting condition for matching of electron and hole pockets is the most favorable for the density wave formation. It was shown that appearance of the "hot spots" at high symmetry points of the Fermi surface implies non-Fermi liquid behavior exhibiting signatures of power laws with universal anomalous dimensions. This results in a non-Fermi liquid metal with a vanishing quasiparticle weight and quasiparticle dispersion with anomalous algebraic momentum dependencies near the "hot spots" \cite{Debbeler}.

We therefore argue that the linear temperature dependence of the magnetic susceptibility in the YCOB single-crystal samples \cite{Zheng}, that is in contrast with the Pauli paramagnetism or Curie-Weiss law expected in conventional metals and magnets, can be explained similarly as in iron-pnictides \cite{Zhang} as a consequence of strong antiferromagnetic correlations within the material. In other words linear temperature dependence of the magnetic susceptibility indicates that the transition to commensurate SDW is not induced only by the Fermi surface nesting \cite{Dong, Mazin} and cannot be described within an itinerant electron approach without taking into account antiferromagnetic SDW correlations. In accordance with the Mermin-Wagner theorem two-dimensional kagome Heisenberg system that we studied can not order at non-zero temperature. However there is a temperature range at lower temperatures where strong local antiferromagnetic SDW correlations exist without established global SDW order. 

The paper is organized as follows. In Sec. \ref{sec:NQS_SR} we review formalism of the symmetry enhanced NQS an SR free energy optimization method. In Sec. \ref{sec:TWBC} we explain flux insertion procedure and twisted boundary conditions. Fermionic parton representation for spin operators is briefly reviewed in Sec. \ref{sec:partons}. We present our results for the ground state properties in Sec. \ref{sec:C_SDW} and compare the results with the results obtained previously with other methods in Sec. \ref{sec:Comparison}. In the last section, Sec. \ref{sec:Conclusions} we draw our conclusions and discuss future research directions. 

\section{Symmetry enhanced NQS ans\"atze and SR method for the free energy optimization}
\label{sec:NQS_SR}
\begin{figure}[b!]
\includegraphics[width=\columnwidth]{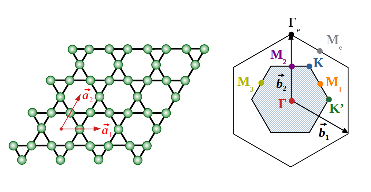}
\caption{\label{Fig:BZ}Kagome lattice first and extended Brillouin zones. Two primitive vectors of the lattice unit cell containing three inequivalent sites are denoted by $\vec{a}_1$ and $\vec{a}_2$ and two primitive basis vectors of the reciprocal lattice by $\vec{b}_1$ and $\vec{b}_2$. $\Gamma$, $K$ and $M$ are high symmetry points in the Brillouin zone. 
}
\end{figure}

A quantum spin-1/2 kagome antiferromagnet in an applied field is described by the following Heisenberg Hamiltonian:
\begin{equation}\label{eq:H_AFM_h}
\hat{H}=J\sum_{\langle i,j \rangle} \hat{\vec{S}}_i\cdot\hat{\vec{S}}_j-h\sum_{i=1}^{N}\hat{S}_i^z,
\end{equation}
where $J > 0$ is the exchange coupling, $\langle i,j \rangle $  denotes nearest-neighboring pairs of the kagome lattice sites and $N$ is the total number of sites. Henceforth, we fix $J=1$, which sets the energy scale for the problem. All other Hamiltonian parameters, such as $h$, are expressed in units of $J$. The spin 1/2  operators $\hat {\vec{S}}_i = \frac{1}{2}\left(\hat{\sigma}^x,\hat{\sigma}^y,\hat{\sigma}_z\right)$ where $\hat{\sigma}_i$ are Pauli matrices with $i \in \{x,y,z\}$. For the kagome lattice with periodic boundary conditions, the lattice Brillouin zone (BZ) for the set of crystal momenta related to the lattice translations is illustrated in FIG. \ref{Fig:BZ}. The lattice unit cell contains three inequivalent sites and positions of the lattice sites are $\vec{R}_{n_1,n_2,k}=n_1\vec{a}_1+n_2\vec{a}_2+\vec{r}_k$ where $\vec{a}_1=(1,0)$ and $\vec{a}_2=(1/2,\sqrt{3}/2)$ are two primitive vectors of the lattice unit cell and $\vec{r}_k$ positions of the sites within the unit cell: $\vec{r}_1=(1/2,0)$, $\vec{r}_2=(1/4,\sqrt{3}/4)$ and $\vec{r}_3=(3/4, \sqrt{3}/4)$. Two primitive basis vectors of the reciprocal lattice are $\vec{b}_1=2\pi(1,-1/\sqrt{3})$ and $\vec{b}_2=4\pi(0,1/\sqrt{3})$.  

To characterize properties of the ground state at the $1/9$ magnetization plateau, through simulations of finite sized lattices, we first employ recently introduced machine learning (ML) approach based on symmetry enhanced GCNNs \cite{Duric,Roth}. The approach has so far shown great promise and potential for studying various complex interacting quantum many-body systems \cite{Duric, Roth, Vecsei} since it combines powerful GCNN NQS variational wave-function ans\"atze and advanced optimization techniques. Additional important advantage of the approach is its suitability for the implementation on graphics processing units (GPUs) that allow multiple parallel computations resulting in significant computational speedup. Our numerical calculations were performed using such computational speedup on GPUs and powerful 
NetKet \cite{Carleo, Vicentini}, JAX \cite{Frostig}, FLAX \cite{Heek} and OPTAX \cite{Hessel} ML libraries. 

The space group that describes symmetries of the kagome lattice is composed of lattice translations, associated with crystal momentum, and dihedral $D(6)$ point group symmetries. If the space group is denoted by $G$ then each symmetry operator $\hat{g} \in G$ commutes with the Hamiltonian (\ref{eq:H_AFM_h}), 
\begin{equation}\label{eq:symmetry}
\left[\hat{H},\hat{g}\right]=0,
\end{equation}
and the eigenstates of the Hamiltonian can be represented as irreducible representations (irreps) of the symmetry group $G$. Within group theory representations are eigenfunctions of the symmetry group operators $\hat{g}$. While reducible representations are eigenfunctions only for a subset of symmetry group operators, irreducible representations are eigenfunctions of all operators $\hat{g}\in G$. Then each irreducible representation has different set of eigenvalues called characters and denoted here by $\chi_g$. 

\begin{figure}[t!]
\includegraphics[width=\columnwidth]{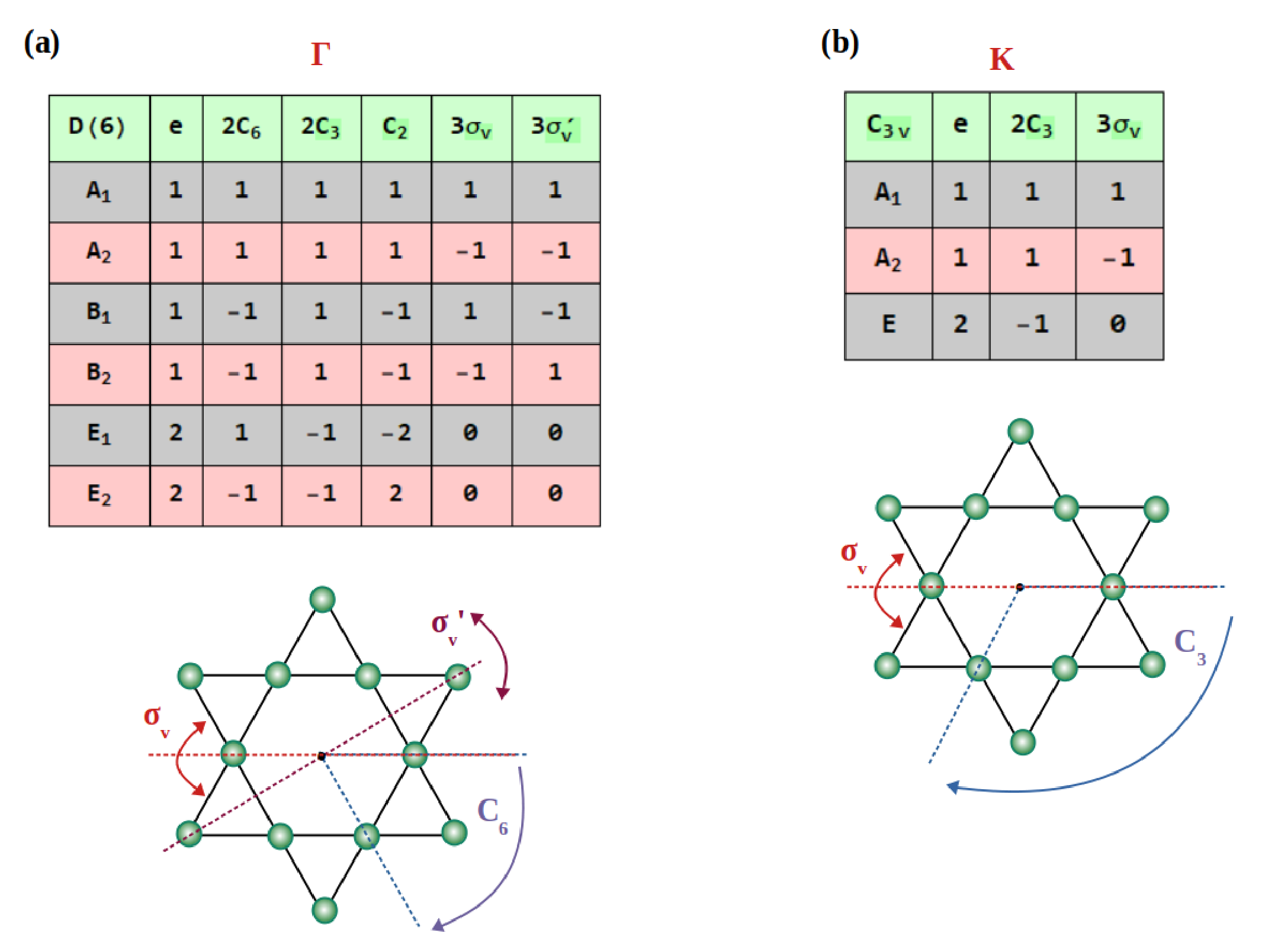}
\caption{\label{Fig:Little_group_gamma_K}Group elements and characters of the kagome lattice little group for the crystal momentum (a) $\vec{q}=(0,0)$ ($\Gamma$ high symmetry point in the kagome lattice BZ) and (b) $\vec{q}=(2\pi/3,2\pi/\sqrt{3})$ ($K$ high symmetry point in the kagome lattice BZ). 
}
\end{figure}

If an irrep is non-degenerate, that is if it consists of only one eigenfunction, characters for irreps are $+1$ or $-1$. Such irreps correspond to non-degenerate eigenstates of the Hamiltonian. On the other hand, degenerate eigenstates correspond to degenerate irreps with characters that can differ from $\pm 1$. The space group irreps can be constructed and described in terms of symmetry related crystal momenta called star and a subgroup of the point group called little group that leaves symmetry related crystal momenta invariant. Group elements and characters of the kagome lattice little group for the crystal momenta that correspond to $\Gamma$ and $K$ points in the BZ illustrated in FIG. \ref{Fig:BZ} are shown in FIG. \ref{Fig:Little_group_gamma_K}. We point out that for both $\Gamma$ and $K$ high symmetry points little group has degenerate irreps (denoted by $E$, $E_1$ and $E_2$) which is of great importance for appearance of the SDW ground state described in the following sections. On the contrary little group for the crystal momentum that corresponds to $M$ high symmetry point does not have degenerate irreps \cite{Duric}.

For interacting spins on the lattice neural network ans\"atze for the Hamiltonian eigenstates, named in literature neural network quantum states (NQS), in general associate a complex number $\psi(\vec{\sigma};\vec{\alpha})$ 
with each spin basis configuration $|\vec\sigma\rangle=|\sigma_1,...,\sigma_{N_s}\rangle$:
\begin{equation}\label{eq:NQS}
|\psi\rangle= \sum_{\vec{\sigma}}\psi(\vec{\sigma};\vec{\alpha})|\vec{\sigma}\rangle
\end{equation}
where $\vec{\alpha}$ denotes the network parameters that can be found by neural network training with a suitably chosen optimization algorithm that minimizes relevant loss function. As it will be described further in this section in our calculations the loss function corresponds to free energy and the chosen optimization algorithm is stochastic reconfiguration (SR) algorithm. Within such machine learning (ML) algorithm spin configurations $|\vec{\sigma}\rangle$ are then the network inputs and the complex coefficients $\psi(\vec{\sigma};\vec{\alpha})$ the network outputs.

GCNN NQS ans\"atze take into account all space-group symmetries, lattice translations and point group symmetries such as rotations and reflections, for the relevant nonabelian symmetry group $G$. GCNN deep neural network architectures can be in general composed of an arbitrary number of layers $N_l$. The first layer is an embedding layer that generates feature maps from the input data:
\begin{equation}\label{eq:GCNN_l1}
 \mathbf{f}_g^1=\Gamma\left(\sum_{\vec{r}}\mathbf{K}\left(\;\hat{g}^{-1}\vec{r}\;\right)\sigma\left(\;\vec{r}\;\right)\right)
 \end{equation}
 where $\mathbf{K}$ denotes a set of learnable kernels or filters that extract local features from the input and $\Gamma$ is a non-linear activation function taken here to be scaled exponential linear unit (SELU) activation function applied separately to the real and imaginary components \cite{Duric,Roth}. 
The output of the embedding layer is further passed to group convolutional layers denoted by $i=2,..,N_l$ with feature to feature convolutions:
\begin{equation}\label{eq:GCNN_W}
\mathbf{f}_g^{i+1}=\Gamma\left(\sum_{\hat{h}\in G}\mathbf{W}^i\left(\hat{h}^{-1}\hat{g}\right)\mathbf{f}_h^i\right)
\end{equation}
where $\mathbf{W}$ is a set of learnable convolutional kernels. Each layer can have $N_f$ such complex-valued feature maps composed of two real-valued feature maps.

To obtain the wavefunction coefficients $\psi(\vec{\sigma};\vec{\alpha})$ in Eq. (\ref{eq:NQS}) the final output layer exponentiates complex-valued feature maps, calculates sum of these exponentiated values, and projects the sum on a particular irrep of the symmetry group $G$ that corresponds to the symmetry properties of the ansatz wavefuntion:
\begin{equation}\label{eq:ansatz_GCNN_a}
\psi(\vec{\sigma})=\sum_{\hat{g}\in G}\chi_g^*\sum_{n=1}^{N_f}\exp\left(f^{N_l}_{n,g}\right).
\end{equation}
The nonlinear activation function before the output is identity function. 

To obtain the lowest energy states in each symmetry sector the GCNN ans\"atze are further optimized with a suitably chosen optimization scheme. In our study we implement variational Monte Carlo (VMC) scheme combined with the natural gradient descent (NGD) method. The NGD method introduces a metric matrix $M$ that takes into account generally non-flat metric
of the parameter manifold unlike the standard gradient descent (GD) method that presumes the flat metric on the parameter manifold with distance equivalent to the Euclidean distance. Within the NGD method the parameter updates are then defined with the equation
\begin{equation}\label{eq:NGD_a}
\vec{\alpha}_{t+1}=\vec{\alpha}_t-\eta M^{-1} \vec{\nabla}_{\vec{\alpha}}\mathcal{L},
\end{equation}
where $\mathcal{L}$ is a suitably chosen loss function that the VMC scheme aims to minimize and $\eta$ is the learning rate in the optimization (neural network training) algorithm.

In our VMC study NGD method corresponds to the to the stochastic reconfiguration method (SR) \cite{Sorella1,Sorella2,Rende,Chen} where the Hilbert-space distance $d$ between two unnormalized wavefunctions $|\phi\rangle$ and $|\phi'\rangle$ is given by the Fubini-Study metrics:
\begin{equation}\label{eq:FS_metrics_a}
\mbox{d}_{FS}=\arccos\sqrt{\frac{\langle\phi'|\phi\rangle\langle\phi|\phi'\rangle}{\langle\phi'|\phi'\rangle\langle \phi|\phi \rangle}}.
\end{equation}
The SR parameter updates (\ref{eq:NGD_a}) are obtained by applying imaginary time evolution operator $e^{-\Delta \tau \mathcal{L}}\approx 1 -\Delta \tau \mathcal{L}$ to the ansatz wavefunction at imaginary time $\tau$:
\begin{equation}\label{eq:imaginary_t_evolution_a}
|\psi'\left(\vec{\alpha}\left(\tau\right)\right)\rangle=\left(1 -\Delta \tau \mathcal{L}\right)|\psi\left(\vec{\alpha}\left(\tau\right)\right)\rangle,
\end{equation}
where $\Delta\tau$ is a small imaginary time step that corresponds to the learning rate $\eta$ in the Eq. (\ref{eq:NGD_a}). 

After each imaginary time step $\Delta\tau$ the wavefunction $|\psi'\left(\vec{\alpha}\left(\tau\right)\right)\rangle$ is projected to the variational subspace by minimizing $d_{FS}^2$ with $|\phi\rangle=|\psi\left(\vec{\alpha}\left(\tau\right)\right)\rangle+\Delta\tau\partial_\tau\psi\left(\vec{\alpha}\left(\tau\right)\right)$ and $|\phi'\rangle=|\psi'\left(\vec{\alpha}\left(\tau\right)\right)\rangle$ in the Eq. (\ref{eq:FS_metrics_a}) to obtain a new set of parameters $\vec{\alpha}_{t+1}\equiv\vec{\alpha}\left(\tau+\Delta\tau\right)$:
\begin{equation}\label{eq:SR_update_a}
\vec{\alpha}_{t+1}=\vec{\alpha}_t-\Delta\tau G^{-1} \vec{\nabla}_{\vec{\alpha}}\mathcal{L}.
\end{equation}
The metric matrix $M$ then corresponds to the quantum geometric tensor (QGT) \cite{Duric,Roth}. To stabilize calculations and achieve reliable convergence  the $G$ matrix is regularized by a small diagonal shift $\epsilon$. 

In our calculations we use kernel (minSR) formulation of SR (NGD) \cite{Rende,Chen} which leads to exactly the same parameter updates as the standard SR formulation:
\begin{equation}\label{eq:Min_SR}
\delta\vec{\alpha}\equiv\vec{\alpha}_{t+1}-\vec{\alpha}_t =\Delta \tau X (X^TX+\epsilon \mathbb{I}_{2M})^{-1}\vec{f},
\end{equation}
where $X$ is the concatenation of the real and imaginary part of the centered Jacobian and the vector $\vec{f}$ is the concatenation of the real and imaginary part of the centered local free energy. If the number of parameters is denoted by $P$ and the number of samples by $M$ Jacobian matrix dimension is $P\times 2M$. Advantage of minSR formulation is that computation of the parameter updates requires inversion of a $M\times M$ matrix instead of inversion of $P\times P$ matrix necessary in the standard SR approach. The minSR formulation is therefore very useful when $P\gg M$ which is typically the case for deep neural network architectures. Since the number of parameters within the GCNN and VMC approach increases with the increase of the system size minSR is particularly advantageous for calculations for larger system sizes. 

The most common choice for the loss function $\mathcal{L}$ is variational energy. Recent studies however showed that adding a pseudoentropy reward term with an effective temperature T to the energy loss function encourages more even sampling of the Hilbert space and helps with instabilities at the initial stages of the training \cite{Duric, Roth}. We therefore minimize the free energy loss function:
\begin{equation}\label{eq:free_energy}
\mathcal{L}_F\equiv F=E-TS,
\end{equation}
where the entropy is defined as
\begin{equation}\label{eq:entropy}
 S=-\sum_{\vec{\sigma}}\frac{|\psi\left(\vec{\sigma},\vec{\alpha}\right)|^2}{\sum_{\vec{\sigma'}}|\psi\left(\vec{\sigma'},\vec{\alpha}\right)|^2}\log \frac{|\psi\left(\vec{\sigma},\vec{\alpha}\right)|^2}{\sum_{\vec{\sigma'}}|\psi\left(\vec{\sigma'},\vec{\alpha}\right)|^2},
\end{equation}
 and $\psi\left(\vec{\sigma},\vec{\alpha}\right)$ is an unnormalized neural-network ansatz wavefunction. Within this approach the effective temperature is set to some initial $T=T_0$ at the start of the training . The temperature is then gradually lowered to zero as the training proceeds according to a schedule function $T_n$ in the nth training step. In our calculations $T_n=T_0\cdot e^{-\lambda n}$ with $T_0=0.5$ and $\lambda=0.02$.

\section{Flux insertion - twisted boundary conditions}
\label{sec:TWBC}
To further elucidate properties of the ground state we apply the flux insertion method \cite{Peierls,Hazra,Sachdev,Oshikawa,Paramekanti}. The effects of the inserted magnetic flux in quantum many body systems can be simulated by introducing twisted boundary conditions (TWBC) \cite{Lin,Zawadzki,Watanabe}. The approach is crucial for definition of topological invariants for interacting many-body quantum systems and characterization of strongly correlated topological phases of matter such as fractional quantum Hall states and fractional Chern insulators \cite{Laughlin,Niu,Kudo, Watanabe2,Oshikawa2}. Moreover the method was applied to reveal signatures of Dirac cones in density matrix renormalization group calculations (DMRG) for kagome and triangular antiferromagnets \cite{He3,Zhu,He4,Hu}. Additionally, it was recently shown that topological invariants defined through the TWBC are equivalent to topological invariants defined through the center-of-mass momentum states in multi-particle systems \cite{Lin}.

\begin{figure}[b!]
\includegraphics[width=\columnwidth]{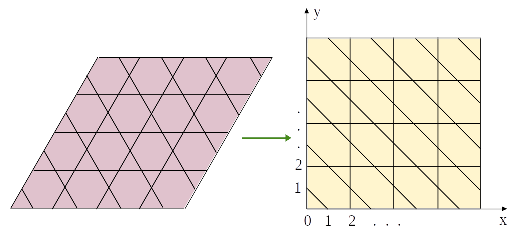}
\caption{\label{Fig:xy_Htwpg} $x(\vec{r}_i)$ and $y(\vec{r}_i)$ coordinates in the Peierls substitution described by equation (\ref{eq:Peierls substition}).}
\end{figure}
Flux insertion can be achieved by a Peierls substitution \cite{Peierls,Hazra,Sachdev,Oshikawa,Paramekanti,He3,Zhu,He4,Hu}:
\begin{equation}\label{eq:Peierls substition}
\hat{S}_{\vec{r}_i}^+\hat{S}_{\vec{r}_j}^- \rightarrow e^{i\left[\frac{\Phi_x}{2\cdot L_x}(x(\vec{r}_j)-x(\vec{r}_i))+\frac{\Phi_y}{2\cdot L_y}(y(\vec{r}_j)-y(\vec{r}_i))\right]} \hat{S}_{\vec{r}_i}^+\hat{S}_{\vec{r}_j}^-
\end{equation}
\begin{figure}[t!]
\includegraphics[width=\columnwidth]{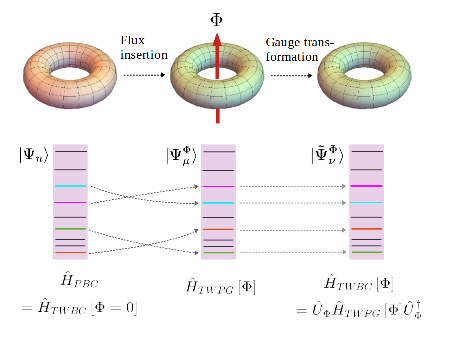}
\caption{\label{Fig:Flux_insertion_method}Flux insertion method: Insertion of unit flux quantum through a torus cycle corresponds to a unitary transformation of the original Hamiltonian
and therefore does not change many-body energy spectrum. After flux insertion the system is described by the Hamiltonian $\hat{H}_{TWPG}$ referred to as twisted Hamiltonian with periodic gauge that can be obtained from the original Hamiltonian by Peierls substitution. The unit flux insertion in general changes crystal momentum and the final state after flux insertion can be different than the initial state. The flux insertion also in general breaks lattice point group symmetries. Since the eigenstates of the original Hamiltonian $\hat{H}_{PBC}$ are connected to the eigenstates of $\hat{H}_{TWPG}$ by a unitary transformation there is however still one-to-one correspondence between the eigenstates. Since the flux can be distributed in various ways it is equivalent to consider energy spectrum of the Hamiltonian with twisted boundary conditions, $\hat{H}_{TWBC}$, related to the Hamiltonian $\hat{H}_{TWPG}$ by a unitary transformation. Under the boundary gauge for $\hat{H}_{TWBC}$ translation symmetry is broken. There is however still one-to-one correspondence between the eigenstates in the periodic gauge and the eigenstates in the boundary gauge since they are related by a unitary transformation $|\psi\rangle_{TWBC}=\hat{U}_{\Phi}|\psi\rangle_{TWPG}$.}
\end{figure}
where $\hat{S}_{\vec{r}_i}^{\pm}=\hat{S}_{\vec{r}_i}^x\pm i\hat{S}_{\vec{r}_i}^{y}$, $L_x\times L_y\times 3$ is the kagome lattice, that with periodic boundary conditions (PBC) forms a torus, and $\Phi_x$ and $\Phi_y$ denote fluxes inserted through two cycles of the torus. For lattice sites at $\vec{r}_i$ coordinates $(x(\vec{r}_i), y(\vec{r}_i))$ are depicted in FIG. \ref{Fig:xy_Htwpg} . The twisted Hamiltonian with periodic gauge obtained by the Peierls substitution (\ref{eq:Peierls substition}) from the original Hamiltonian (\ref{eq:H_AFM_h}), $H_{TWPB}(\Phi_x,\Phi_y)$, in general depends on the flux showing Aharonov-Bohm effect which is absent when $\Phi_x$ and $\Phi_y$ correspond to the unit flux quantum. Inserting one flux quantum through a cycle of the torus, for example the cycle denoted by $y$, corresponds to fluxes $\Phi_x=0$ and $\Phi_y=2\pi$. FIG. \ref{Fig:Flux_insertion_method} illustrates the unit flux quantum insertion through one of the cycles of the torus.

The magnetic flux can be distributed in various ways. It is therefore equivalent to consider energy spectrum of the linearly transformed Hamiltonian:
\begin{equation}\label{eq:Htwbc}
\hat{H}_{TWBC}(\vec{\Phi})=U_{\vec{\Phi}}\hat{H}_{TWPG}(\vec{\Phi})U_{\vec{\Phi}}^{-1},
\end{equation}
where 
$\vec{\Theta}=(\Theta_x,\Theta_y)$ and the transformation $U_{\vec{\Phi}}$:
\begin{equation}\label{U_twbc}
U_{\vec{\Phi}}=e^{i\sum_{\vec{r}} \left [ \frac{\Phi_x}{2\cdot L_x}\cdot x(\vec{r}) +\frac{\Phi_y}{2\cdot L_y}\cdot y(\vec{r})\right]\cdot \hat{S}_{\vec{r}}^z}.
\end{equation}
The linearly transformed Hamiltonian $\hat{H}_{TWBC}$ is equivalent to the original Hamiltonian (\ref{eq:H_AFM_h}), however with twisted boundary conditions:
\begin{eqnarray}\label{eq:TWBC}
&\hat{S}^z_{\vec{r}_{L_x}}=\hat{S}^z_{\vec{r}_0}, \;\; \hat{S}_{\vec{r}_{L_x}}^{\pm}=e^{\mp \cdot i \cdot \phi_x}\hat{S}_{\vec{r}_0}^{\pm}, \\ 
&\hat{S}^z_{\vec{r}_{L_y}}=\hat{S}^z_{\vec{r}_0}, \;\; \hat{S}_{\vec{r}_{L_y}}^{\pm}=e^{\mp \cdot i \cdot \phi_y}\hat{S}_{\vec{r}_0}^{\pm}, \nonumber
\end{eqnarray}
where $\vec{r}_0=(0,0)$, $\vec{r}_{L_x}=L_x\cdot \vec{a}_1$ and  $\vec{r}_{L_y}=L_y\cdot \vec{a}_2$. We note that $\hat{H}_{TWBC}$ is a periodic function of $\Phi_x$ and $\Phi_y$:
\begin{eqnarray}\label{eq:TWBC_2pi}
\hat{H}_{TWBC}(\Phi_x,\Phi_y)&=&\hat{H}_{TWBC}(\Phi_x+2\pi,\Phi_y) \\ \nonumber
&=&\hat{H}_{TWBC}(\Phi_x,\Phi_y+2\pi)\\
&=&\hat{H}_{TWBC}(\Phi_x+2\pi,\Phi_y+2\pi),\nonumber
\end{eqnarray}
while the Hamiltonian with periodic gauge $H_{TWPB}$ does not have periodicity of $2\pi$ with respect to the twist angles $\Phi_x$ and $\Phi_y$. 

Flux insertion in general breaks lattice point group symmetries. In addition flux insertion changes canonical momentum $-i\vec{\nabla}$ since canonical momentum is not gauge invariant. However, kinetic momentum, equivalent to covariant derivative $-i\vec{\nabla}-\vec{A}$ with $\vec{A}$ being the vector potential that corresponds to the inserted flux, is gauge invariant. Gauge transformation due to flux insertion therefore in general changes crystal momentum of an eigenstate \cite{Hazra,Oshikawa,Paramekanti,Lin}. Namely, if $n$ flux quanta are inserted through, for example, cycle $y$ of the torus with $\Phi_y=2\pi n$, the crystal momentum changes as:
\begin{equation}\label{eq:Kp}
\vec{q}\rightarrow \vec{q}'=\vec{q}-\Delta\vec{q}
\end{equation}
where $\Delta\vec{q}$:
\begin{equation}\label{eq:delta_K}
\Delta\vec{q}=\frac{n\langle \sum_i \hat{S}_i^z\rangle}{2L_y}\vec{b}_2=\frac{nL_x}{12}\vec{b}_2,
\end{equation}
since $\langle \sum_i \hat{S}_i^z\rangle=L_x\times L_y/6$ for the $3\times L_x\times L_y$ kagome lattice at the $1/9$ magnetization plateau. We find the lowest energy states for the twist angle $\Phi_y=\pi L_y$ (corresponding to $n=L_y/2$ flux quanta) and original wavevectors $\vec{q}$ at the $\Gamma$ and $K$ points in the kagome lattice BZ. For the $6\times 6\times 3$ lattic,e the twist angle effectively changes original crystal momentum for $\vec{b_2}/2$ since adding reciprocal lattice primitive basis vector to the original crystal momentum does not change the crystal momentum due to the lattice periodicity. For the $12\times12\times 3$ lattice, the original wavevector remains unchanged.

As it will be explained in the following section this result reveals presence of Dirac cones in the energy spectrum of the underlying fermionic spinons. Namely, similarly as for Dirac spin liquid \cite{He3} fermionic spinons see an additional emergent gauge flux $\phi_e$. For symmetry reasons $\phi_e$ can be $0$ or $\pi$ corresponding to two topological sectors that are degenerate in the thermodynamic limit. An emergent flux $\phi_e=\pi$ corresponds to antiperiodic boundary condition for spinons and $\phi_e=0$ to periodic boundary condition for spinons. For a finite size system with $L_y$ of the form $L_y=4k+2$ the ground state energy is lower for antiperiodic boundary condition \cite{He3}. Since spinons see half of the inserted flux $\Phi_y/2$ we therefore find the lowest energy state for the $6\times 6\times 3$ lattice when the inserted flux $\Phi_y=6\pi$ corresponding to antiperiodic boundary condition for spinons. We note that the flux $\Phi_y=6\pi$ instead of the flux $\Phi_y=2\pi$ (as in the case for Dirac spin liquid \cite{He3}) is necessary because spinons have an effective charge $1/3$ due to uniform $2\pi/3$ flux in each kagome lattice unit cell \cite{He, Zheng}. The $12\times12\times3$ lattice (with $L_y$ of the form $L_y = 4k$), however, behaves differently and the lowest energy state that we find corresponds to the periodic boundary conditions for spinons with zero emergent gauge flux. We note that both topological sectors are degenerate in the thermodynamic limit. Similar difference between $4k$ and $4k+2$ cylinders was found in DMRG calculations \cite{He3}.

Additionally our results indicate possibility of emergent particle-hole symmetry since the twist angle $\Phi_y=L_y\pi$ is necessary for the twisted Hamiltonian in periodic gauge to remain invariant under particle-hole transformation \cite{Zawadzki}. Such emergent particle-hole symmetry was found for example for electrons in a half-filled Landau level that can be described as Dirac fermions coupled to an emergent gauge field \cite{Son, Mulligan, Nguyen,Wang,Bhatt}.

We also note that the insertion of a unit flux quantum does not change many-body energy spectrum since it corresponds to a unitary transformation of the original Hamiltonian. Since the eigenstates of the original Hamiltonian $H_{PBC}$ and the twisted Hamiltonian with periodic gauge $H_{TWPB}$ are connected by the unitary transformation there is one-to-one correspondence between the eigenstates although flux insertion in general changes momentum and breaks lattice point group symmetries. As it will be described in the following section inserting $n=L_i/2$ $i\in\{x,y\}$ flux quanta through one of the cycles of the torus can greatly simplify numerical calculations ultimately leading to much more accurate results for the ground state properties of the system.  

\section{Fermionic parton representation for spin operators and emergent gauge field}\label{sec:partons}
In this section we briefly review fermionic parton representation for the spin-1/2 operators and an effective theory of fermionic spinons coupled to an emergent gauge field that can help explain results presented in the following section. The spin-1/2 operators can be represented in terms of fermionic spinons:
\begin{equation}\label{eq:spinons}
\hat{\vec{S}}_i=\frac{1}{2}\hat{\psi}_i^{\dagger}\hat{\vec{\sigma}}\hat{\psi}_i, 
\end{equation}
where $\hat{\psi}_i=(\hat{f}_{i,\uparrow},\hat{f}_{i,\downarrow} )^T$, $\hat{f}_{i,\alpha} \; \alpha \in \{\uparrow,\downarrow\}$ are fermionic spinon operators, and $\vec{\sigma}=(\hat{\sigma}^x,\hat{\sigma}^y,\hat{\sigma}^z)$ are Pauli matrices. The representation (\ref{eq:spinons}) is used in most of the standard VMC frameworks \cite{He}. The mapping between Hilbert spaces for spins and fermionic spinons is exact, provided fermionic spinons adhere to the constraint of no double occupancy:
\begin{equation}\label{eq:constraint}
\hat{n}_{i,\uparrow} + \hat{n}_{i,\downarrow}=1 \;\; \forall i,
\end{equation}
where $\hat{n}_{i,\alpha}=\hat{f}_{i,\alpha}^\dagger\hat{f}_{i,\alpha} \; \alpha \in\{\uparrow,\downarrow\}$ denotes number operator for fermions $\alpha$.

When the Hamiltonian (\ref{eq:H_AFM_h}) is rewritten in terms of spinons and decoupled into a quadratic mean-field Hamiltonian the resulting Hamiltonian is:
\begin{equation}\label{eq:H_mf}
H_{MF}=\sum_{\langle i,j \rangle}(\hat{\psi}_i^{\dagger}\hat{U}_ij\hat{\psi}_j + \hat{\psi}_j^{\dagger}\hat{U_ij}^{\dagger}\hat{\psi}_i) -\sum_i \mu \hat{\psi}_i^{\dagger}\hat{\sigma}^z\hat{\psi}_i,
\end{equation}
where $\hat{U}_{ij}$ and $\mu$ form variational parameters. We note that $H_{MF}$ is the zeroth order mean-field Hamiltonian that does not take into account gauge fluctuations of the gauge field on the links of the lattice. Namely, the representations (\ref{eq:spinons}) has SU(2) gauge redundancy \cite{Wen2} and the mean-field decoupling of the spin interactions therefore has to include SU(2) gauge field on the links of the lattice
\begin{eqnarray}\label{eq:gauge_su2}
\hat{\psi}_i &\rightarrow& \hat{W}(i)\hat{\psi}_i\\
\hat{U}_{ij}&=&\hat{W}(i)U_{ij}\hat{W}(j)\nonumber
\end{eqnarray}
Saddle points of the mean field theory can however break this $SU(2)$ symmetry down to $U(1)$ or $Z_2$. The zeroth order mean field theory is not always sufficient to describe properties of the system and gauge fluctuations around the mean-field solution have to be included into consideration. For example $U(1)$ gauge fluctuation around the mean-field solution $\hat{\bar{U}}_{ij}$ can be written as $\hat{U}_{ij}=\hat{\bar{U}}_{ij}e^{\hat{a}_{ij}^z\hat{\sigma}^z}$.

Variational calculations based on Gutzwiller projected mean field ansatz wavefunctions $|\psi\rangle =P_G|\psi\rangle_{MF}$, where $P_G=\prod_{i=1}^N (1-\hat{n}_{i,\uparrow}\hat{n}_{i,\downarrow})$, found that the ground state of the Hamiltonian (\ref{eq:H_AFM_h}) corresponds to a $Z_3$ spin liquid \cite{He}. Renormalization group (RG) calculations for composite Fermi liquids \cite{Jian} however indicate that when gauge fluctuations are included underlying composite Fermi liquid shows $2k_F$ instability resulting in a density wave ground state. Including gauge fluctuations leads to an additional interaction term in the total effective action $S=S_f+S_a+S_{int}$ \cite{Jian}:
\begin{equation}\label{eq:}
S_{int}\propto \int d^3xa(x)\hat{\psi}^{\dagger}(x)\hat{\psi}(x)
\end{equation}
where $a(x)$ is the emergent gauge field and $S_{int}$ describes interaction between fermions and emergent gauge bosons. As it will be demonstrated in the following section our results clearly demonstrate such instability and appearance of a CSDW.

We further briefly explain effects of the flux insertion on fermionic spinons mentioned in the previous section. From the Eq. (\ref{eq:spinons}) it is evident that if original spins see an inserted flux of $\Phi_y$ then fermionic spinons see an inserted flux $s_{\sigma}\Phi_y/2$ where $s_{\sigma}=+1$ or $-1$ for $\uparrow$ and $\downarrow$ spinons, respectively. For $n$ inserted flux quanta with $\Phi_y=2n\pi $ change in the crystal momentum $\vec{q}_{\sigma}$ for spinons is then \cite{Lin}:
\begin{equation}\label{eq:Kp_spinons}
\vec{q}_{\sigma} \rightarrow \vec{q}_{\sigma} -\delta \vec{q}_{\sigma},
\end{equation}
where $\vec{q}_{\sigma}$ is
\begin{equation}\label{eq:Delta_K_spinons}
\delta \vec{q}_{\sigma}=s_{\sigma}\frac{n N_{\sigma}}{4L_y}\vec{b}_2,
\end{equation}
and $N_{\sigma}$ is the total number of $\sigma$ fermions. The total change of the momentum for spins is then:
\begin{equation}\label{eq:Delta_K_spins_spinons}
\delta\vec{q}=\delta\vec{q}_{\uparrow} -\delta\vec{q}_{\downarrow} =\frac{n}{2L_y}\frac{(N_{\uparrow}-N_{\downarrow})}{2}\vec{b}_2
\end{equation}
Taking into account that $\langle \sum_i \hat{S}_i^z\rangle=(N_\uparrow - N_{\downarrow})/2$, $\delta\vec{q}$ is then described by equation (\ref{eq:delta_K}).

As predicted by recently introduced Dirac spinon model \cite{Zheng}, spinon energy spectrum features both larger Fermi surface and Dirac cones. The results presented in the following section reveal presence of Dirac cones and underlying spinon Fermi surface and also its instability towards density wave formation.

\section{Ground state properties: Chiral spin density wave}
\label{sec:C_SDW}
We started our study by optimizing the free energy loss function (\ref{eq:free_energy}) with GCNN ans\"atze that take into account full dihedral $D(6)$ symmetry group and translation symmetry for the kagome lattices with $N=6\times6\times3=108$ sites and periodic boundary conditions (torus geometry). To study the properties of the ground state at the $1/9$ magnetization plateau, we first attempted to optimize the free energy within the fixed magnetization sector $M_z^{tot}/M_z^s=1/9$, where $M_z^{tot}=6$ is the total magnetization of the system and $M_z^s=(1/2)\cdot N$ is the saturation  magnetization, with sampling performed by exchanging two neighboring spins or two random spins. Optimization within the fixed magnetization sector showed very poor convergence. The constrained optimization stays caught in a local minimum far from the lowest energy values obtained for the ground state energy with other methods. Therefore we optimized the free energy within the unconstrained optimization scheme by taking into account the whole Hilbert space and by sampling with the single spin flip updates. 
\begin{figure}[b!]
\includegraphics[width=\columnwidth]{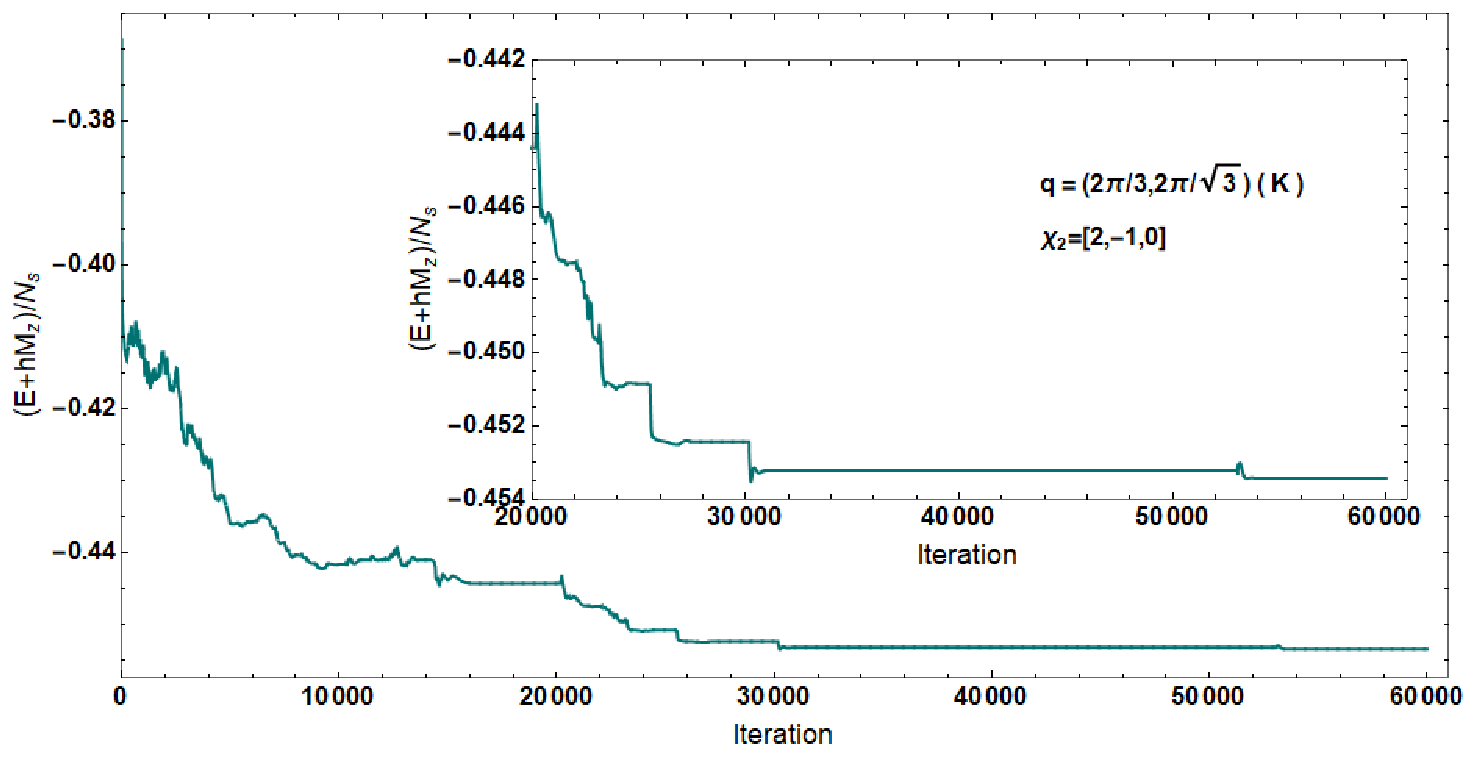}
\caption{\label{Fig:E0_Phi0}Variational energy per site for the GCNN ansatz with crystal momentum $\vec{q}=(2\pi/3,2\pi/\sqrt{3})$ ($K$ point in the kagome lattice BZ) and irrep $E$. The inset shows the data after the simulation has reached fixed average magnetization $1/9$ during pretraining. The lowest found energy per site is $\approx-0.4534$. 
}
\end{figure}
\begin{figure}[t!]
\includegraphics[width=\columnwidth]{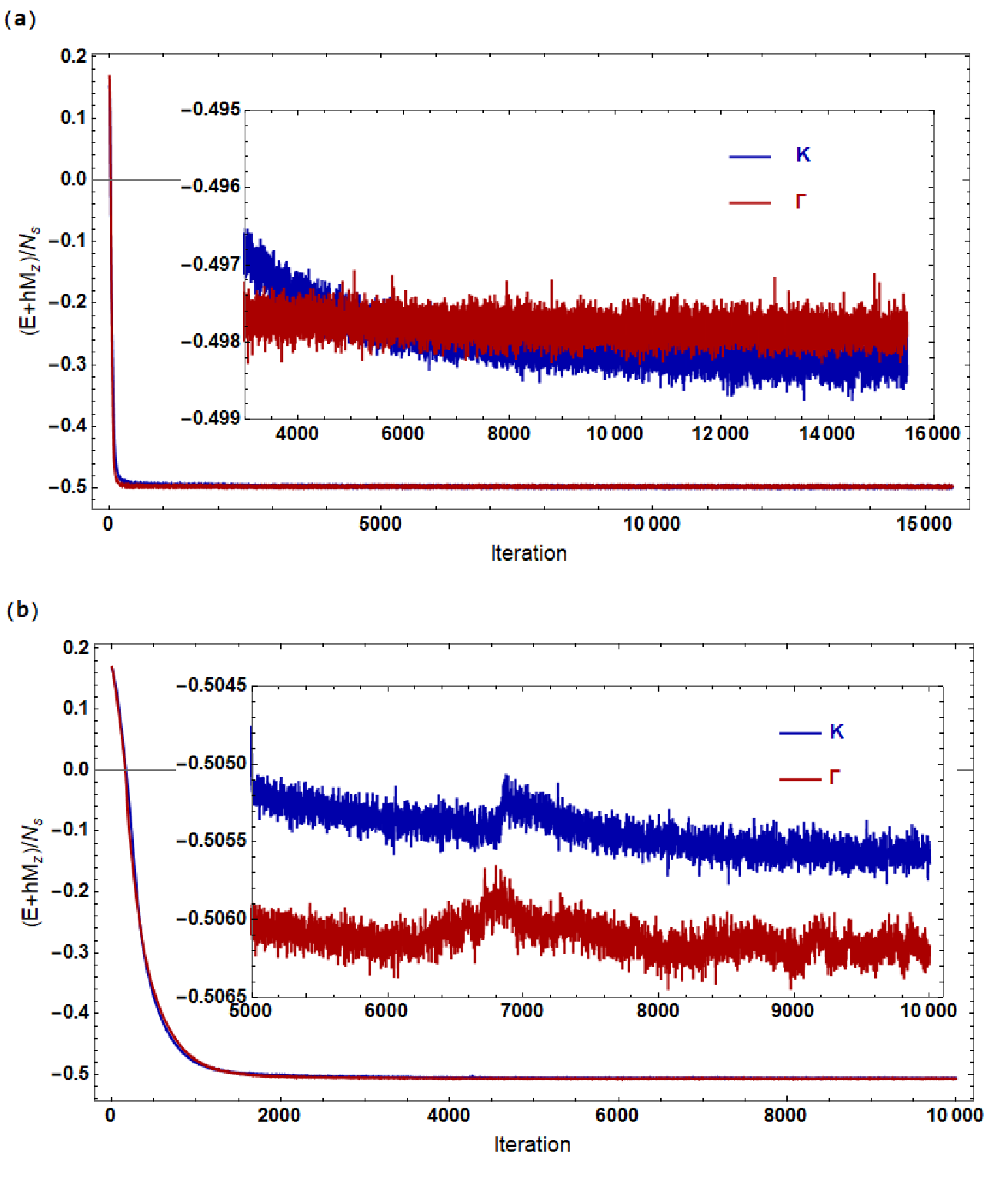}
\caption{\label{Fig:E0}Variational energy per site obtained by the flux insertion method supplemented approach for the lowest energy eigenstates with the original crystal momenta at $\Gamma$ and $K$ points in the kagome lattice BZ corresponding to $\vec{q}=(0,0)$ and $\vec{q}=(2\pi/3,2\pi/\sqrt{3})$, respectively. The variational energy is calculated for $N=6\times 6\times 3=108$ (panel (a)) and $N=12\times 12\times 3=432$  (panel (b)) sites kagome lattices with periodic boundary conditions. The inserted flux is  $(\Phi_x,\Phi_y)=(0,\pi L_y)$. The results are calculated within the fixed magnetization sector corresponding to the $1/9$ plateau and the energy due to the Zeeman field term in the Hamiltonian (\ref{eq:H_AFM_h}) is subtracted.  For $N=108$ lattice sites, the original crystal momenta are shifted by $\vec{b}_2/2$ upon flux insertion where $\vec{b}_2$ is a primitive basis vector of the reciprocal lattice.}
\end{figure}

We find that the unconstrained scheme provides improved results due to the improved ability of the scheme to escape local minima. The final results are shown in FIG. \ref{Fig:E0_Phi0}. The lowest energy eigenstates are found at the crystal momentum $\vec{q}=(2\pi/3,2\pi/\sqrt{3})$ corresponding to the $K$ point in the kagome lattice Brillouin zone shown in FIG. \ref{Fig:BZ}. For both constrained and unconstrained scheme NQS architecture contains $N_l=4$ layers with $N_f=4$ features in each layer. Learning rate $\eta$ in the optimization algorithm is set to $0.02-0.05$ and Monte Carlo (MC) estimates are calculated with $2^{12}$ samples. Local cluster for which elements of the embedding kernels $K$ (Eq. (\ref{eq:GCNN_l1})) are nonzero is chosen to be a hexagon of 6 sites around origin for lattice sites \cite{Duric}. The first step was to find optimized eigenstates corresponding to three irreps shown in the table (b) in FIG. \ref{Fig:Little_group_gamma_K}. The calculations were performed for several values of the Zeeman field $h$ in the Hamiltonian (\ref{eq:H_AFM_h}), with the  $h$ values chosen in the vicinity of the center of the $1/9$ magnetization plateau found by DMRG \cite{Nishimoto} and iPEPS calculations \cite{Fang}. We note that since the total magnetization operator $\hat{M}_z^{tot} =\sum_{i=1}^N \hat{S}_i^z$ commutes with the Hamiltonian (\ref{eq:H_AFM_h}), $[\hat{H},\hat{M_z^{tot}}]=0$, the lowest energy eigenstate is the same anywhere on the plateau with constant fixed average magnetization. We also point out that the convergence is significantly improved by applying transfer learning for various $h$ values and irreps. Our calculations found that the plateau magnetization value ( $M_z^{tot}=6$ for $N=108$ lattice sites) is reached for the lowest energy states corresponding to all three irreps demonstrating existence of several low energy states at the plateau. The lowest energy eigenstate is finally found for the irrep denoted by $E$ in the table (b) in  Fig. \ref{Fig:Little_group_gamma_K}.

The irrep $E$ is doubly degenerate. Degeneracy of the ground state further indicates possibility for the spontaneous breaking of the lattice point group symmetries. This will be confirmed by additional calculations using the flux insertion method presented later in this section. We note that convergence to the ground state energy is slow and tens of thousand iterations are necessary to identify irrep corresponding to the ground state. The results for the lowest energy state are shown in FIG. \ref{Fig:E0_Phi0}. The best estimate of the ground state obtained from the simulations is an energy eigenstate with  momentum K and symmetries described by the irrep E. This state has energy $E_0\approx -0.4534$ per site which is lower than the best estimates obtained in previous studies.~\cite{He} However, as evident from FIG. \ref{Fig:E0_Phi0}, the optimization procedure encounters several local minima where the simulation remains stuck for thousands of iterations before finally succeeding in escaping the minima (even after pretraining of all three irreps with transfer learning at various values of the Zeeman field at the plateau). After around $10^6$ iterations, we notice degradation of data quality and overfitting that causes slow increase of the loss function. Described optimization scheme therefore does not conclusively rule out the existence of lower energy eigenstates. 

\begin{figure}[t!]
\includegraphics[width=\columnwidth]{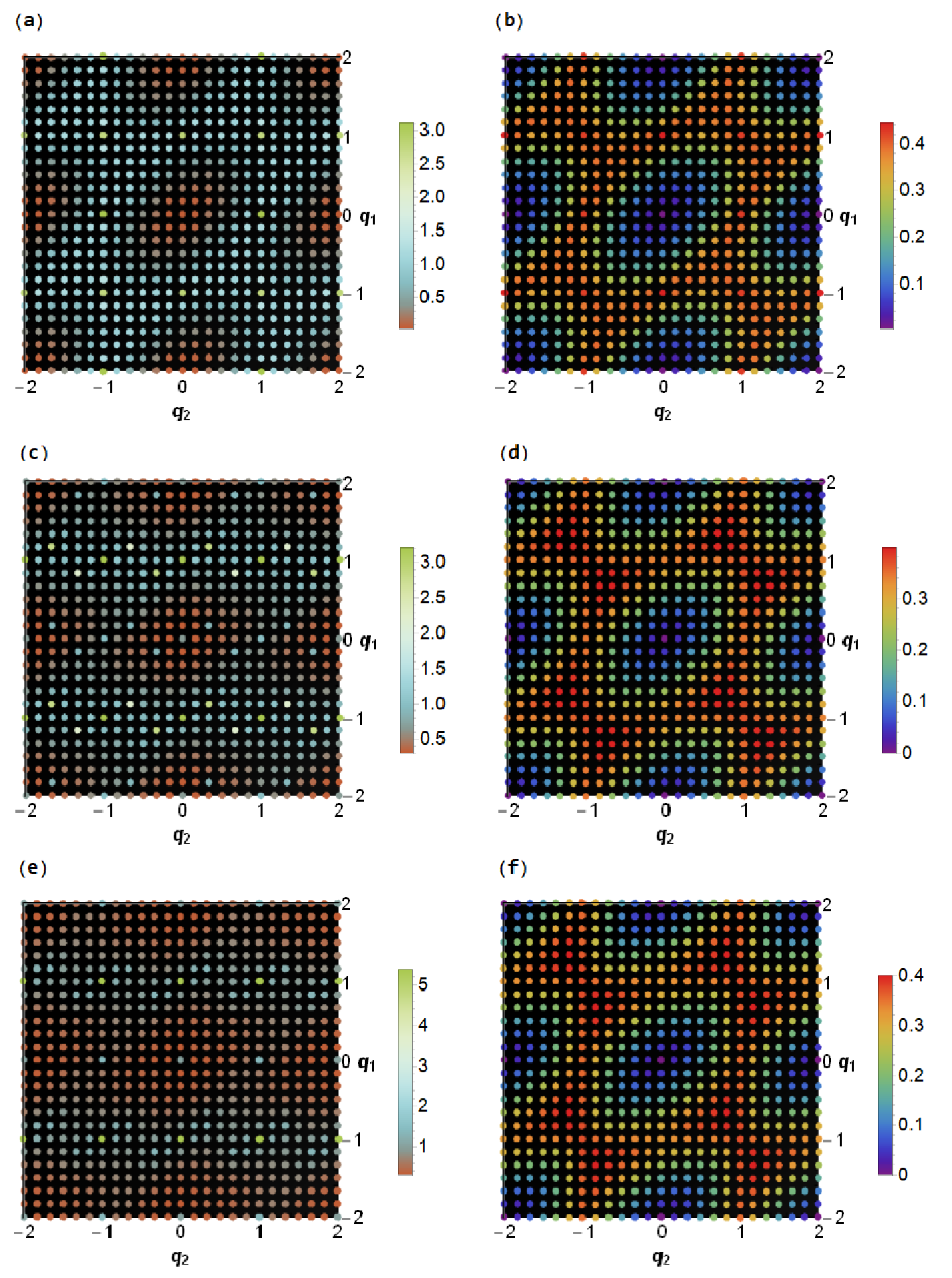}
\caption{\label{Fig:Sf_6x6x3} Static spin structure factors $S_f(\vec{q})$ (panels (a), (c) and (e)) and $S_f^z(\vec{q})$ (panel (b), (d) and (f)) for the lowest energy states of the $N=6\times 6\times 3=108$ sites kagome lattice with periodic boundary conditions. Panels (a) and (b) are results from simulations without inserted flux and for the lowest energy state with crystal momentum $\vec{q}=(2\pi/3,2\pi/\sqrt{3})$ ($K$ point in the kagome lattice BZ). Panels (c) - (f) present results of simulations with inserted $(\Phi_x,\Phi_y)=(0,6\pi)$ flux - (c) and (d) are for the lowest energy state with the original crystal momentum $\vec{q}=(2\pi/3,2\pi/\sqrt{3})$ and panels (e) and (f) for the lowest energy state with the original crystal momentum $\vec{q}=(0,0)$ ($\Gamma$ point in the kagome lattice BZ). For $N=108$ lattice sites, the original crystal momenta are shifted by $\vec{b}_2/2$ upon flux insertion where $\vec{b}_2$ is a primitive basis vector of the reciprocal lattice. Structure factors demonstrate the appearance of the spinon Fermi surface (contours of blue hexagons in $S_f^z(\vec{q})$ in panels (b), (d) and (f) correspond to $2k_F$) and an instability of the Fermi surface towards formation of a SDW visible as pronounced peaks in $S_f(\vec{q})$ (panels (a), (c) and (e)). The structure factors are shown as functions of the wave vector $\vec{q}=(q_1,q_2)$, where components $q_1=n_1/L_1$ and $q_2=n_2/L_2$ are in units of the primitive basis vectors $\vec{b}_1$ and $\vec{b}_2$ shown in Fig. \ref{Fig:BZ} with $L_1=L_2=6$ and $n_i$ integers. 
}
\end{figure}

To improve stability and convergence of the results we further supplement our calculations with the flux insertion method described in Sec. \ref{sec:TWBC}. NQS GCNN architectures for the flux insertion supplemented calculations are chosen to have $N_l=6$ layers and $N_f=6$ features. Local cluster for which embedding kernels (Eq. (\ref{eq:GCNN_l1})) in the GCNN ans\"atze are nonzero is again chosen to be hexagon of 6 sites around the origin for the kagome lattice sites \cite{Duric}.  The lowest energy eigenstates are found for the wavefunctions at crystal momenta $\vec{q}=(0,0)$ and $\vec{q}=(2\pi/3,2\pi/\sqrt{3})$ corresponding to the $\Gamma$ and $K$ high symmetry points of the kagome lattice BZ (Fig. \ref{Fig:BZ}) and variational energies per site for the lowest energy states and kagome lattices with $N=6\times6\times 3=108$ and $N=12\times12\times 3=432$ sites are shown in Fig. \ref{Fig:E0}. The results in Fig. \ref{Fig:E0} are calculated with the inserted flux $(\Phi_x,\Phi_y)=(0,\pi L_y)$ and  within the fixed magnetization sector corresponding to the $1/9$ plateau ($M_z^{tot}=6$ and $M_z^{tot}=24$ for $N=108$ and $N=432$ sites, respectively). 

Fig. \ref{Fig:E0} clearly demonstrates improved stability, convergence and accuracy of the results. The lowest energy eigenvalues are: $E(K,N=108)\approx -0.4984$, $E(\Gamma,N=108)\approx -0.4978$, $E(K,N=432)\approx -0.5056$ and $E(\Gamma,N=432)\approx -0.5062$. For the $N=108$ sites  lattice, eigenstate with the crystal momentum corresponding to the $K$ point in the kagome lattice BZ has slightly lower energy while for the $N=432$ site lattice, the eigenstate with zero crystal momentum at $\Gamma$ point has lowest energy. For $N=108$ lattice sites, the original crystal momenta are shifted by $\vec{b}_2$ upon flux insertion where $\vec{b}_2/2$ is a primitive basis vector of the reciprocal lattice. We argue that two states are degenerate in the thermodynamic limit revealing nontrivial underlying topology \cite{Gioia}. 

Flux insertion in general breaks lattice point group symmetries. The GCNN ans\"atze for the eigenstates of the twisted Hamiltonian in periodic gauge therefore retain only lattice translation symmetry. Although the lattice point group symmetries are broken there is one-to-one correspondence between the eigenstates of the original Hamiltonian and the twisted Hamiltonian in periodic gauge for the inserted flux $(\Phi_x,\Phi_y)=(0,\pi L_y)$ since the two Hamiltonians are connected by the unitary transformation defined in Eq.~(\ref{U_twbc}). It is important to note that the lowest energy eigenstate obtained in the standard GCNN simulations has significant overlap with that obtained using flux insertion, even though the energy is slightly higher.  Hence, based on the preceding arguments, we can impose the symmetry properties of the former on the latter. Accordingly,  we argue that the found lowest energy states correspond to degenerate states described by irreps $E$, both at $\Gamma$ and $K$ points (Fig.~\ref{Fig:Little_group_gamma_K}). For degenerate irreps, characters in the character table refer to the transformation of all components and correspond to trace of the transformation matrix for degenerate states vector. The ground state for twisted Hamiltonian in periodic gauge can therefore correspond to nonsymmetrized combination of degenerate states. 
 
Having obtained an estimate for the ground state energy and an individuation of the corresponding ground state wavefunction in terms of symmetries, we proceed to determine the nature of the ground state by calculating several physical observables. We start with the static spin structure factor. 
The total static spin structure factor is calculated as:
\begin{equation}\label{eq:Ssf}
S_f(\vec{q}) = \frac{1}{N}\sum_{i,j}e^{i\vec{q}\cdot(\vec{r}_i-\vec{r}_j)}\left(\langle \vec{S}_i\cdot\vec{S}_j\rangle -\langle \vec{S}_i\rangle\cdot\langle \vec{S}_j\rangle\right).
\end{equation}
We have also calculated the longitudinal component of the static spin structure factor that corresponds to $\langle S_i^z \cdot S_j^z\rangle$ correlations:
\begin{equation}\label{eq:Ssf_z}
S_f^z(\vec{q}) = \frac{1}{N}\sum_{i,j}e^{i\vec{q}\cdot(\vec{r}_i-\vec{r}_j)}\left(\langle S^z_i\cdot S_j^z\rangle -\langle S_i^z\rangle\cdot\langle S_j^z\rangle\right).
\end{equation}

\begin{figure}[t!]
\includegraphics[width=\columnwidth]{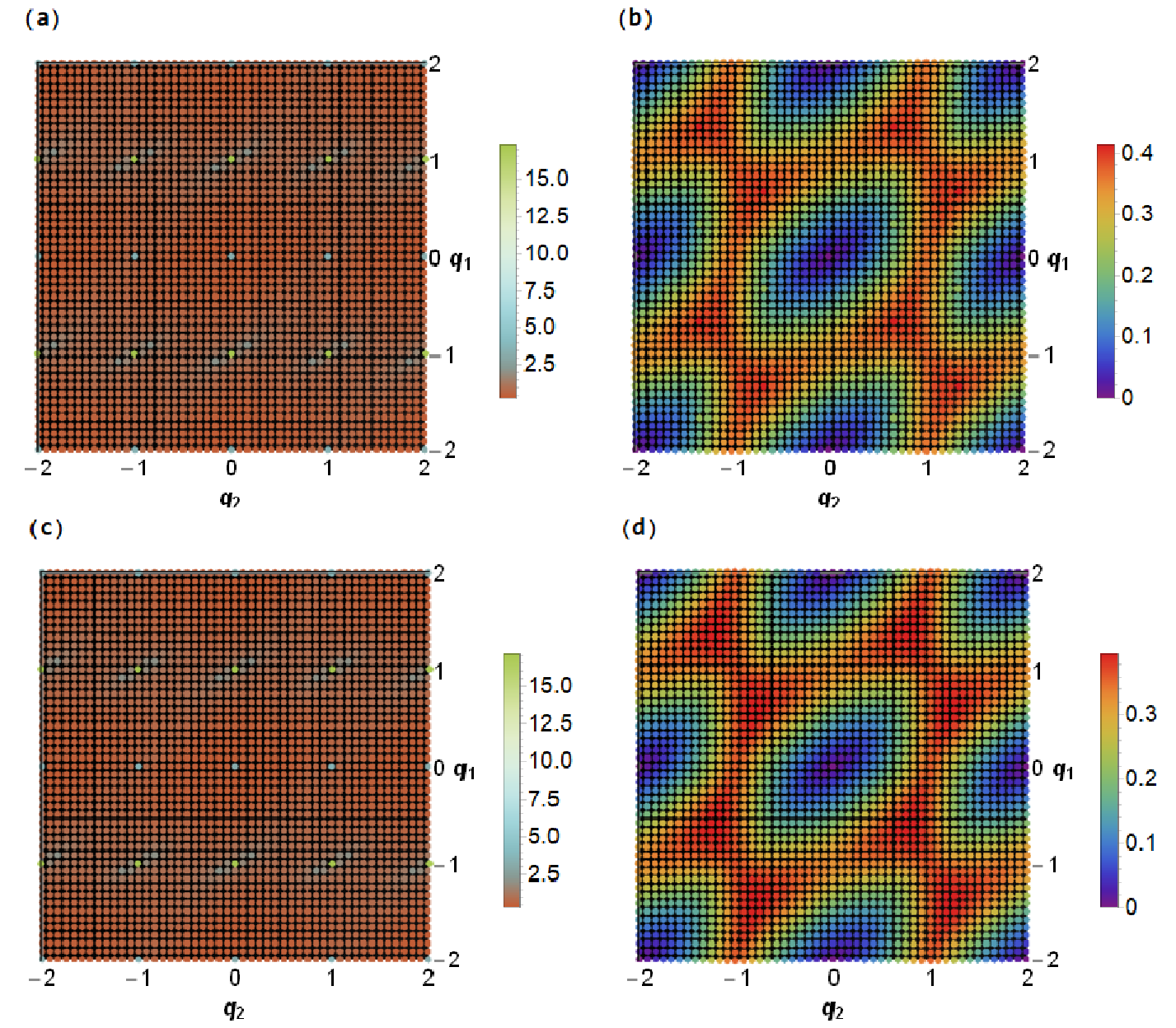}
\caption{\label{Fig:Sf_12x12x3}  Static spin structure factors $S_f(\vec{q})$ (panels (a) and (c)) and $S_f^z(\vec{q})$ (panels (b) and (d)) calculated for the lowest energy states of $N=12\times 12\times 3=432$ sites kagome lattice with periodic boundary conditions and inserted $(\Phi_x,\Phi_y)=(0,12\pi)$ flux. Panels (a) and (b) show results for the lowest energy state with crystal momentum $\vec{q}=(2\pi/3,2\pi/\sqrt{3})$ ($K$ point in the kagome lattice BZ) and panels (c) and (d) for the lowest energy state with crystal momentum $\vec{q}=(0,0)$ ($\Gamma$ point in the kagome lattice BZ). The structure factors are shown as functions of the wave vector $\vec{q}=(q_1,q_2)$, where components $q_1=n_1/L_1$ and $q_2=n_2/L_2$ are in units of the primitive basis vectors $\vec{b}_1$ and $\vec{b}_2$ shown in Fig. \ref{Fig:BZ} with $L_1=L_2=12$ and $n_i$ integers. 
}
\end{figure}
The calculated static spin structure factors $S_f(\vec{q})$  and $S_f^z(\vec{q})$ for the lowest energy states of $N=6\times 6\times 3=108$ and $N=12\times 12\times 3=432$ sites kagome lattices are shown in Fig. \ref{Fig:Sf_6x6x3} and Fig. \ref{Fig:Sf_12x12x3}, respectively. The structure factors are shown as functions of the wave vector $\vec{q}=(q_1,q_2)$, where components $q_1=n_1/L_1$ and $q_2=n_2/L_2$ are in units of the primitive basis vectors $\vec{b}_1$ and $\vec{b}_2$ shown in Fig. \ref{Fig:BZ} with $n_i$ integers for finite lattices with $N=L_1\times L_2\times 3$ sites. Panels (a) and (b) in Fig. \ref{Fig:Sf_6x6x3} show calculated  $S_f(\vec{q})$ and $S_f^z(\vec{q})$ for the standard GCNN simulations whereas the rest are for simulations with inserted $(\Phi_x,\Phi_y)=(0,L_y\pi)$ flux with $L_y=L_2$. Panels (c) and (d) in Fig. \ref{Fig:Sf_6x6x3} and Fig. \ref{Fig:Sf_12x12x3} present results for the lowest energy state with the original crystal momentum $\vec{q}=(2\pi/3,2\pi/\sqrt{3})$ (K point in the kagome lattice BZ) and panels (e) and (f) for the lowest energy state with the original crystal momentum $\vec{q}=(0,0)$ ($\Gamma$ point in the kagome lattice BZ). MC estimates for the spin structure factors are calculated from 2$^{16}$ samples.  $S_f(\vec{q})$  and $S_f^z(\vec{q})$ for the lowest energy eigenstate obtained from the standard GCNN simulation exhibit qualitatively similar features as those for the true ground state obtained from the flux insertion method, underscoring the fact that it is adiabatically connected to the true ground state.  

\begin{figure}[b!]
\includegraphics[width=\columnwidth]{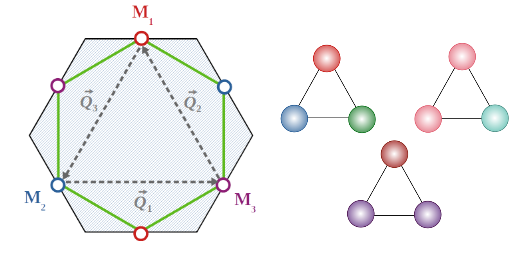}
\caption{\label{Fig:Fermi_surface}Illustration of the found spinon Fermi surface (green hexagon), Van Hove singularities of the kagome lattice (M points in the BZ) and possible CSDW patterns resulting from the Fermi surface instability (right panel). Fermi surface nesting wavevectors are denoted by $\vec{Q}_i$.
}
\end{figure}
Spin structure factors can reveal scattering features close to the Fermi surface \cite{Jian, Altshuler,Patel,Holder} with the Fermi wavevector $k_F$ and also density-wave instabilities that cause sharp peaks in the structure factors at the wavevector corresponding to the density-wave wavevector \cite{He,Xu,Sheng}.  Structure factors clearly demonstrate appearance of the spinon Fermi surface (contours of blue hexagons in $S_f^z(\vec{q})$ in panels (b), (d) and (f) correspond to $2k_F$) and an instability of the Fermi surface towards formation of a SDW is visible as pronounced peaks in $S_f(\vec{q})$ in panels (a), (c) and (e). We note that sharp peaks due to SDW instability do not appear in $S_f^z(\vec{q})$ showing that SDW instability corresponds to appearance of the $XY$ condensate \cite{Sachdev2}. Figures also show noticeable smaller peak in structure factors at $\Gamma$ point in the kagome lattice BZ indicating existence of Dirac point at $\Gamma$ and Dirac-like excitations in its vicinity.

Namely, the spin structure factors can be rewritten in terms of the scattering functions for spinons \cite{Patel2} defined as:
\begin{equation}\label{eq:scattering_function}
F(\vec{k}) \sim \sum_{\vec{q}}A(\vec{q},\omega=0)A(\vec{q}+\vec{k},\omega=0)
\end{equation}
that capture momentum transfer $\vec{k}$ (from $\vec{q}$ to $\vec{q}+\vec{k}$) scattering across the Fermi surface with energy transfer $\omega=0$, where the spinon spectral function $A(\vec{k},\omega)=\delta(\omega -\epsilon_F^{S}(\vec{k}))$. The spinon Fermi surface found from the $2k_F$ surface revealed in the spin structure factors is illustrated in Fig. \ref{Fig:Fermi_surface}. Particularly important feature of the Fermi surface is the Fermi surface nesting that occurs when one portion of the Fermi surface shifted by a nesting wavevector $\vec{Q}_i$ overlaps with another portion of the Fermi surface. The Fermi surface nesting is beneficial for interaction induced instabilities that cause appearance of spin or charge density waves. Additional important feature of the fermionic spinon bands are Van Hove singularities. In general Van Hove singularities arise from the extreme or saddle points in the fermionic spinon bands and for the kagome lattice they appear at the M points in the BZ. Such points are characterized by divergent density of states and are often referred to as critical points of the BZ. When Van Hove singularities are near Fermi level and the Fermi surface is nested by a nesting wavevector $\vec{Q}_i$ numerous instabilities, and in particular interaction driven CSDW instability, can occur causing Fermi surface reconstruction.

We note that although Fermi surface nesting itself in some cases can explain CDW/SDW instability (in particular in one-dimensional systems) in many cases strong correlations and interactions play crucial role in a density wave formation. Although it is not possible to disentangle contribution of the interactions and correlations from the contribution of the Fermi surface nesting in our spin structure factors calculations we argue that for kagome antiferromagnets, interactions and correlations are evidently of great importance. As explained in the introduction a linear temperature dependence of the magnetic susceptibility found in the recent experiments on YCOB single-crystal \cite{Zheng} appears as a consquence of the SDW antiferromagnetic correlations similarly as in iron-pnictides \cite{Zhang}. Our results are therefore consistent with recent experimental findings.

\begin{figure}[t!]
\includegraphics[width=\columnwidth]{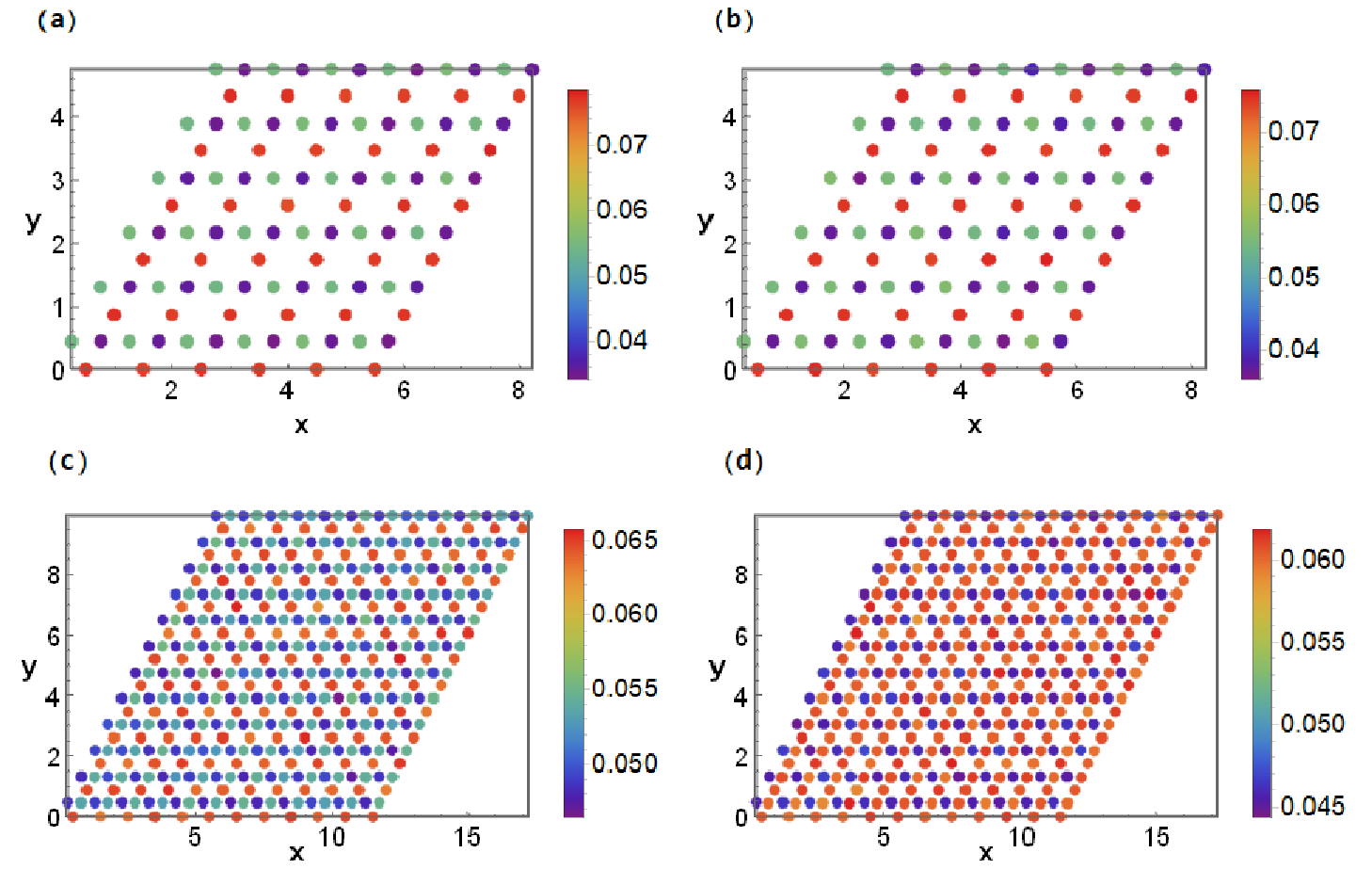}
\caption{\label{Fig:Sz}The expectation values for the local magnetization $\langle S_i^z\rangle$ for the lowest energy states with the original crystal momentum at $K$ (panels (a) and (c)) and $\Gamma$ (panels (b) and (d)) points in the kagome lattice BZ and for $N=6\times 6\times 3=108$ (panels (a) and (b)) and $N=12\times 12\times 3=432$ (panels (c) and (d)) sites kagome lattices  with periodic boundary conditions. The calculated  $\langle S_i^z\rangle$ values clearly demonstrate that the lowest energy states have $1\times 1$ CSDW order. We also note that $\langle S_i^z\rangle$ is evidently invariant with respect to any gauge transformation of the form as in Eq. (\ref{U_twbc}). For $N=108$ lattice sites, the original crystal momenta are shifted by $\vec{b}_2/2$ upon flux insertion where $\vec{b}_2$ is a primitive basis vector of the reciprocal lattice.
}
\end{figure}
To probe the topological character of the lowest energy states obtained in the simulations, we have calculated the chiral order parameter defined as:
\begin{equation}\label{eq:Ch_op}
|\chi|=\frac{1}{2}\left(|\chi_\bigtriangleup| +|\chi_\bigtriangledown|\right)
\end{equation}
where 
\begin{equation}\label{eq:Ch_op_triangle}
\chi_{\bigtriangleup/\bigtriangledown} =\frac{1}{N_{\bigtriangleup/\bigtriangledown}}\sum_{i\in \bigtriangleup/\bigtriangledown} \vec{S}_{i_1} \cdot \left(\vec{S}_{i_2}\times \vec{S}_{i_3}\right)
\end{equation}
with indices 1, 2, and 3 corresponding to the clockwise arranged corners of $\bigtriangleup/\bigtriangledown$ elementary triangles of the kagome lattice, and $N_{\bigtriangleup/\bigtriangledown}$ corresponding to the total number of triangles. Our calculations reveal that the lowest energy states obtained in all the simulations have finite chiral order parameter. For the standard GCNN simulation, the lowest energy state features $|\chi|\approx 0.039$ with $\chi_{\bigtriangledown}\approx0.044$ and $\chi_{\bigtriangleup}\approx0.034$. 
For the simulations with flux insertion, nonzero values for the chiral order parameters (Eqs. (\ref{eq:Ch_op}) and (\ref{eq:Ch_op_triangle}) ) are observed for all lowest energy states: $|\chi|(N=108,\;K)\approx 0.044$ ($\chi_\bigtriangledown\approx -0.004$, $\chi_\bigtriangleup\approx -0.084$), $|\chi|(N=108,\; \Gamma)\approx 0.052$ ($\chi_\bigtriangledown\approx -0.022 $, $\chi_\bigtriangleup\approx -0082$), $|\chi|(N=432, \; K)\approx 0.055$ ($\chi_\bigtriangledown\approx -0.025$, $\chi_\bigtriangleup\approx -0.085$), and $|\chi|(N=432, \;\Gamma)\approx 0.058$ ($\chi_\bigtriangledown\approx -0.029$, $\chi_\bigtriangleup\approx -0.086$), where MC estimated for chiral order parameters are calculated from $2^{20}$ samples.
Finite chiral order parameters indicate time reversal symmetry breaking and appearance of a chiral SDW (CSDW) with nontrivial underlying topology. As it will be shown further in this section true ground state found by supplementing our calculations with the flux insertion method is also a CSDW with visible modulation of $\langle\vec{S}_i^z\rangle$ within the kagome unit cell (Fig. \ref{Fig:Sz} (a) and (b)). Appearance of charge-density-wave (CDW), SDW and CSDW patterns was found so far in several kagome metals \cite{Wen,OBrien,Nishimoto2, Ferrari,Sachdev2, Teng,Kim,Neupert,Jiang,Lee2,Christensen,Ruegg,Holbaek,Liu,Park,Huang}.

Taken together, the results for the low energy states illustrate $2k_F$ scattering features caused by scattering close to the spinon Fermi surface illustrated in Fig. \ref{Fig:Fermi_surface} and CSDW instability causing sharp peaks at crystal momentum values corresponding to $2\vec{Q}_i$ where $\vec{Q}_i$ are the nesting wavevectors illustrated in Fig. \ref{Fig:Fermi_surface}.

\begin{figure}[b!]
\includegraphics[width=\columnwidth]{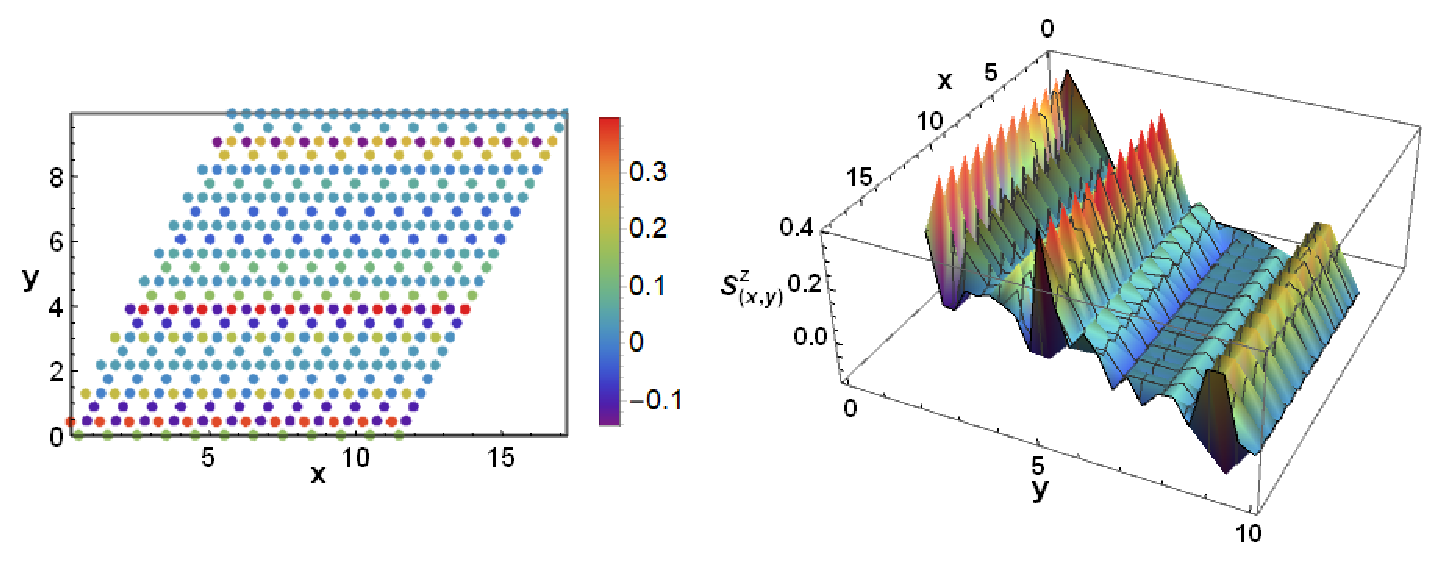}
\caption{\label{Fig:Friedel_oscillations}Generalized Friedel oscillations visible in the average value of the spin z-component, $\langle \hat{S}_i^z\rangle$, for the CSDW ground state for the $N=12\times 12\times 3=432$ site kagome lattice with cylinder geometry. The crystal momentum corresponding to the periodic direction of the cylinder is zero. 
}
\end{figure}
To determine the nature of spin density modulation in the CSDW state, we have calculated the average value of the spin z-component, $\langle S_i^z\rangle$, that is trivially gauge invariant. MC estimates are calculated from 2$^{20}$ MC samples. The results for the lowest energy states (at the K and $\Gamma$ points of the kagome lattice BZ) obtained from simulations with flux insertion are shown in Fig.~\ref{Fig:Sz} for systems sizes $N=6\times 6 \times 3 = 108$ and $N=2\times 12\times 3 = 432$ sites. The results confirm 1 × 1 density wave ordering pattern. The $1\times 1$ CSDW states maintain original translation symmetry of the kagome lattice with spin modulation only within the kagome lattice unit cell. Analogous density wave patterns have previously been found in the phase diagram of the kagome-lattice extended Hubbard model at the van Hove filling ($N_{electrons}/N_{sites}=5/6$) \cite{Ferrari} and interacting spinless fermions at $1/3$ filling \cite{Wen}. The CSDW is characterized by a triangle rule, namely each unit cell has one site with $\langle S_i^z\rangle < 0.5 $ and two sites with $\langle S_i^z\rangle > 0.5$. Additionally all CSDW states characterized by the triangle rule are gapless \cite{Ferrari}.

Further confirmation of the existence of the spinon Fermi surface and its instability to CSDW formation is appearance of generalized Friedel oscillations due to the open boundary for the ground state on cylinder geometry \cite{White,Dora, Zhao,Lin2,Zhang2}. Friedel oscillations in a metal or semiconductor are caused by a defect (for example an impurity) in the Fermi gas or Fermi liquid that causes regular modulation of the fermion density around impurity. Namely, scattering of the electrons close to the Fermi suface by a local disturbance and their waves interference results in the formation of regular modulations in the electron charge and spin density with amplitudes decreasing with distance from the disturbace. An open boudary acts equivalently as an impurity and causes Friedel oscillations of the fermion density close to the boundary.  

As explained in Sec. \ref{sec:partons}, $\langle\hat{S}_i^z\rangle$ can be mapped to fermionic spinon density:
\begin{equation}\label{eq:Sz_Friedel_o}
\langle \hat{S}_i^z\rangle =\frac{1}{2}(\hat{n}_{i,\uparrow}-\hat{n}_{i,\downarrow}),
\end{equation}
where $\hat{n}_i^{\sigma}$ denotes the number of $\sigma \in \{\uparrow,\downarrow\}$ spinons at a lattice site $i$, and the spinons adhere to the constraint of no double occupancy $\hat{n}_{i,\uparrow}+\hat{n}_{i,\downarrow}=1$. Oscillations in spinon density therefore correspond to oscillations in $\langle \hat{S}_i^z\rangle$:
\begin{equation}\label{eq:ni_Friedel_o}
\langle \hat{S}_i^z\rangle \sim \hat{n}_i^{\sigma},
\end{equation}
and appearance of Friedel oscillations signals existance of fermionic quasiparticles.

In normal metals, described within non-interacting electrons model, the local electron density at a distance $r$ from the impurity or boundary behaves as \cite{Simion}:
\begin{equation}\label{eq:Friedel_o}
n(r)\approx \frac{\sin(2k_Fr+\phi_d)}{r^d},
\end{equation}
where $d$ is dimension of the system and $\Phi_d$ is a suitable dimensionality-dependent phase. In a density wave phase generalized Friedel oscillations survive beyond the density wave  coherence length, defined as $v_F/\Delta$ where $v_F$ is the Fermi velocity and $\Delta$ denotes density wave order parmater, however with significantely suppressed amplitude \cite{Dora}. Additionally, presence of several Fermi pockets or valleys can lead to multimode Friedel oscillations.

Our calculations for cylinder geometry clearly show generalized multimode Friedel oscillations in the average value of the spin z-component, $\langle S_i^z\rangle$, close to the boundary (that correspond to the Friedel oscillations of the spinon density) as shown in Fig. \ref{Fig:Friedel_oscillations} for the $N=432$ sites kagome lattice. The GCNN ansatz for cylinder geometry contains only translation symmetry corresponding to the periodic direction of the cylinder and we set number of features to $N_f=16$ and number of layers to $N_l=4$. MC estimates in the VMC algorithm for the free energy optimization and $\langle S_i^z\rangle$ MC estimates are calculated with $2^{12}$ and $2^{20}$ samples, respectively. The learning rate used for the free energy optimization is $\eta=0.01$.

\section{CSDW ground state versus $Z_3$ spin liquid and VBS states }
\label{sec:Comparison}
In this section we briefly compare CSDW ground state found by our machine learning approach with the $Z_3$ spin liquid state and two VBS states found by DMRG, iPEPS and VMC calculations based on fermionic parton representation for spin operators. The lowest energy for the $Z_3$ spin liquid state found by DMRG \cite{Nishimoto} and VMC \cite{He} approaches is much higher than the energy of the CSDW state, $E_0(Z_3)/(JN)\approx -0.41178$ \cite{He} (with omitted energy from the Zeeman field) for $N=12\times 12\times 3=432$ lattice sites with periodic boundary conditions. Very similar energy value was found by iPEPS calculations for a $\sqrt{3}\times\sqrt{3}$ VBC state, $E(VBC_{\sqrt{3}\times\sqrt{3}})/(JN) \approx -0.4111$ \cite{Fang,Cheng}. The VBC state has special hourglass $\sqrt{3}\times\sqrt{3}$ structure that breaks the spatial translational invariance with an extended nine-site unit cell, gapless excitations and scaling behaviors of the entanglement entropy and the correlation length that indicate a $c=1$ conformal field theory \cite{Fang}. Recently a different VBC state was found by VMC calculations with a more general variational ansatz based on the resonating valence bond (RVB) picture \cite{Cheng}. This VBC state has $3\times 3$ periodicity, windmill-shaped motif, strong spatial modulation in the local magnetization and significantly lower energy than the $Z_3$ spin liquid state and the $\sqrt{3}\times \sqrt{3}$ VBC state, $E(VBC_{3\times 3})/(JN) \approx -0.4184$ \cite{Cheng}. 

The lowest CSDW energy found by our machine learning approach supplemented by the flux insertion method is $E(CSDW)/(JN)\approx -0.5062$ for the $N=12\times 12\times 3=432$ site lattice with periodic boundary conditions as shown in Fig. \ref{Fig:E0}. We argue that such large discrepancy between the energies obtained with other methods and our machine learning approach stem from the bias of other methods towards particular kinds of states. Namely, matrix product states within DMRG algorithm are biased towards gapped low entangled states and can therefore fail to recognize highly entangled gapless phases like the CSDW found in this work. Similarly various ans\"atze within the VMC calculations based on fermionic parton representation and RVB picture cannot represent all possible relevant low energy states. Previous calculations with other methods therefore find states that correspond to various local minima and not the global energy minimum. 

Additional interesting observation is that all found low energy states show non-trivial spatial modulations in the local magnetization. DMRG calculations find several regular nearly degenerate magnetization density configurations for various finite open kagome lattices at the $1/9$ magnetization plateau \cite{Nishimoto} that are numerically unstable against weak perturbations at the edges of the lattice. Although this numerical instability may indicate that the magnetization density of the 1/9-plateau is structureless in the thermodynamic limit there is no definite proof for such structureless magnetization density in the thermodynamic limit. Similarly both previously found VBC states show regular spatial modulation in the local magnetization with different configurations of the local magnetization components in direction parallel to the Zeeman field. 

\section{Conclusions}
\label{sec:Conclusions}
We have studied ground-state properties of the quantum spin-1/2 kagome antiferromagnet in an applied field at $1/9$ magnetization plateau using recently developed machine learning approach supplemented with flux insertion method. Our method incorporates symmetry enhanced GCNN neural networks that in combination with flux insertion and minSR optimization lead to significant improvement of the achievable results accuracy. Contrary to the results obtained with other methods, that predicted $Z_3$ spin liquid or valence bond crystals exhibiting an hourglass pattern or $3\times3$ pattern with a windmill-shaped motif, our calculations find that the ground state is a CSDW that exhibits $1\times 1$ pattern and has finite chirality and nontrivial underlying topology. In agreement with recent experiments on YCOB single-crystal samples we find evidence for existence of Dirac-like spinons in the low energy spectrum above CSDW ground state and explain experimentally found linear temperature dependence of the uniform  magnetic susceptibility as a consequence of strong CSDW antiferromagnetic correlations. Particularly interesting direction for future research is to examine behavior of the system upon doping which most likely will show significant similarity with behavior found in many high temperature superconducting materials. 

\begin{acknowledgments}
We thank Anders Sandvik and Chisa Hotta for helpful discussions. This work is supported by the Singapore Ministry of Education (MOE) Academic Research Fund
Tier 3 Grant (No. MOE-MOET32023-0003) "Quantum Geometric Advantage". We would also like to acknowledge the NTU High Performance Computing Centre for providing computing resources, facilities, and services that have contributed to this work.
\end{acknowledgments}

\end{document}